\documentclass[pre,twocolumn,9pt]{revtex4}

\newcommand{\dd}{\mathrm{d}}
\newcommand{\ii}{\mathrm{i}}
\newcommand{\erf}{\mathrm{erf}}

\newcommand{\vettoref}{	\boldsymbol{f}}

\newcommand{\vettoreF}{	\boldsymbol{F}}
\newcommand{\vettoreG}{	\boldsymbol{G}}
\newcommand{\vettoreR}{	\boldsymbol{R}}

\newcommand{\vettoreErre}{\boldsymbol{r}}

\newcommand{\vettoreRho}{\boldsymbol{\rho}}

\newcommand{\kversore}{\hat{\boldsymbol{k}}}

\newcommand{\Nabla}{\boldsymbol{\nabla}}

\newcommand{\CalA}{\mathcal{A}}

\usepackage{ae} 
\usepackage{amsmath}
\usepackage{amsfonts}
\usepackage{hyperref}
\usepackage{graphicx}
\usepackage[T1]{fontenc}

\begin{document}

\title{Paraxial Sharp-Edge Diffraction: A General Computational Approach}

\author{Riccardo Borghi}
\affiliation{ Dipartimento di Ingegneria, Universit\`{a} ``Roma Tre'', Via Vito Volterra 62, I-00146 Rome, Italy}


\begin{abstract}
A general reformulation of classical sharp-edge diffraction theory is proposed within paraxial approximation. 
The, not so much known, Poincar\'e vector potential construction is employed directly inside
Fresnel's 2D integral in order for it to be converted  into a single 1D contour integral over the aperture boundary.
Differently from the recently developed paraxial revisitation of BDW's theory, such approach can be applied  
to arbitrary wavefield distributions impinging onto arbitrarily shaped sharp-edge planar apertures. 
A couple of interesting examples of application of the proposed method is presented.
\end{abstract}

\maketitle

\section{Introduction}
\label{Sec:Introduction}

The study of propagation of light  continues to play a central role in Optics and Photonics.
The increasing complexity of modern optical systems poses new challenges to  
light/matter interaction modeling. In particular, sharp-edge diffraction represents a key problem to be tackled 
whenever light is left to pass through a linear optical 
system. Pupils, filters, diaphragms, lenses, unavoidably limit the transverse 
distribution of the incoming wavefield. As a consequence, edge diffraction effects at all relevant 
boundaries have to take into account in order for the field emerging from the exit pupil to be
 adequately characterized (in amplitude and phase).

Sharp-edge diffraction is as old as wave theory of light:  in 1802, Thomas Young  first suggested the 
idea that the rim of an illuminated aperture could act as a secondary light source~\cite{Rubinowicz/1957,Rubinowicz/1965}.
Due to their  physical appeal, Young's ideas received a continuous, growing attention.
But only after the pioneering works by Maggi~\cite{Maggi/1888} and by Rubinowicz~\cite{Rubinowicz/1917}, 
these ideas have definitely found a quantitative formulation in the form of the so-called  BDW 
theory~\cite[Ch. 8]{Born/Wolf/1999}, thought for spherical and/or plane wave diffraction by arbitrarily shaped sharp-edge  
apertures (or obstacles). The first attempt of extending BDW's theory to deal with general impinging 
wavefields was done by Miyamoto and Wolf at the beginning of sixties~\cite{Miyamoto/Wolf/1962a,Miyamoto/Wolf/1962b}. 

Despite its formal elegance, the practical applicability of the BDW/Miyamoto/Wolf theory turned out to not to be as easy as 
it could have been expected, even for apparently simple incoming disturbances, like for instance Gaussian beams.
To develop a more manageable theory, paraxial approximation was then invoked from the beginning. 
In this way, a ``genuinely paraxial'' version of the original Young/Maggi/Rubinowicz theory was proposed in~\cite{Borghi/2015,Borghi/2016}, 
based on some results published in~\cite{Stamnes/1983,Hannay/2000}. This paraxial revisitation of BDW theory soon revealed its predictive potential~\cite{Borghi/2017,Borghi/2018}, 
especially once placed within the context of the so-called \emph{Catastrophe Optics}~\cite{Berry/Upstill/1980,Nye/1999}.
Later on, further generalizations  aimed at dealing with sharp-edge diffraction under Gaussian and Bessel illuminations
have also been proposed in~\cite{Borghi/2019,Borghi/2020} and in~\cite{Borghi/Carosella/2022}, respecively.


The basic issue of sharp-edge diffraction theory is the transformation of the two-dimensional (2D) Kirchhoff integral, 
which implements Huygens' superposition principle, into a \emph{contour} (i.e., 1D) integral over the aperture boundary. 
The  most known mathematical tool to achieve such a conversion is Green's theorem. Gordon~\cite{Gordon/1975} and, indipendently,
Asvestas~\cite{Asvestas/1985a,Asvestas/1985b}, Forbes and Asatryan   later~\cite{Forbes/Asatryan/1998},
emphasized the importance of Poincar\'e's elegant construction of vector potentials~\cite{Yap/2009} as a practical tool for 
achieving surface-to-line conversion of Kirchhoff's integral. Now, the paraxial limit of Kirchhoff's integral is Fresnel's integral.  It would then be natural 
to ask whether it is possible again to invoke paraxial approximation from the beginning, in order for a 
general sharp-edge paraxial diffraction theory to be developed. Up to my knowledge, this does not seem to have yet been proposed. 


The idea is simple: to employ Poincar\'e vector potential construction directly into Fresnel's integral, in order to convert  it 
into a contour integral over the edge. While, in principle, such a conversion turns out to be always possible,
its practical applicability depends on the capacity of analytically solving certain 1D integrals. 
In the present paper it is proved that such conversion is possible for an important subclass of the Laguerre-Gauss beam family. 
This would be enough to explore a virtually infinite variety of different scenarios. 
To give a single example, the \emph{near-field} produced by the sharp-edge diffraction of vortex beams by triangular apertures 
is explored. Similar scenarios have already been analyzed in the past, but limitedly to the \emph{far-field} zone.  

If the conversion to a single contour integral were not analytically achievable, the proposed approach unavoidably leads to a  double integral representation of 
the diffracted wavefield. However, differently from Fresnel's integral, whose domain coincides with the aperture, the new representation turns out to 
always be defined onto a \emph{rectangular} domain, a fact that greatly simplify its numerical computation. To give 
evidence of this, an iconic example will be illustrated:  the focal wavefield distribution produced by  a collimated laser beam 
impinging onto a water droplet lens whose boundary is forced to assume an equilater triangular shape. The simulations are aimed at 
reproducing some of the beautiful experimental results obtained forty years ago by Berry, Nye, and Wright in a seminal paper  
which became part of the Catastrophe Optics manifesto~\cite{Berry/Nye/Wright/1979}. 
All historical considerations aside, this example constitutes an important test to check the practical applicability of the proposed 
approach in rather extreme situations, with Fresnel numbers of the order of thousands, and where the 2D integral can be
numerically evaluated by employing standard Montecarlo integration packages.

\section{Theoretical Analysis}
\label{Sec:AsvestasTheorem}

Consider a scalar disturbance impinging onto a planar, opaque screen having an aperture $\CalA$ delimited by the boundary $\Gamma=\partial \CalA$.
The screen is placed at the plane $z=0$ of a suitable cylindrical reference frame $(\vettoreErre;z)$. On denoting $\psi_0(\vettoreErre)$ the disturbance distribution
at $z=0^-$, the field distribution $\psi$ at the observation point $P\equiv (\vettoreErre;z)$, with $z>0$ is given, within paraxial approximation
and apart from an overall phase factor $\exp(\ii k z)$, by the Fresnel integral
\begin{equation}
\label{Eq:FresnelIntegral}
\begin{array}{l}
\displaystyle
\psi(\boldsymbol{r};U)\,=\,
-\frac{\mathrm{i}U}{2\pi}\,
\int_{\boldsymbol{\mathcal{A}}}\,
\mathrm{d}^2\rho\,
\psi_0(\boldsymbol{\rho})\,
\exp\left(\frac{\mathrm{i}U}{2}\,|\boldsymbol{r}-\boldsymbol{\rho}|^2\right)\,,
\end{array}
\end{equation}
where, in place of $z$, the Fresnel number $U=ka^2/z$ has been introduced. The symbol $a$ denotes a characteristic length of the aperture size (for instance, if 
$\CalA$ were circular, $a$ would coincide with its radius). In this way, only dimensionless  quantities will be involved in the following. 

The whole paraxial scalar diffraction theory is based on Eq.~(\ref{Eq:FresnelIntegral}), which can also be recast in terms of the new integration variable $\vettoreR=\boldsymbol{\rho}+\vettoreErre$ as follows:
\begin{equation}
\label{Eq:FresnelIntegralBis}
\begin{array}{l}
\displaystyle
\psi(\boldsymbol{r};U)\,=\,
-\frac{\mathrm{i}U}{2\pi}\,
\int_{\boldsymbol{\mathcal{A}}}\,
\mathrm{d}^2R\,\,
\psi_0(\vettoreR\,+\,\vettoreErre)\,
\exp\left(\frac{\mathrm{i}U}{2}\,R^2\right)\,,
\end{array}
\end{equation}
and which corresponds to observe the aperture $\CalA$ from a reference frame centred at the observation point $P$.
\begin{figure}[!ht]
\centerline{\includegraphics[width=5cm,angle=-0]{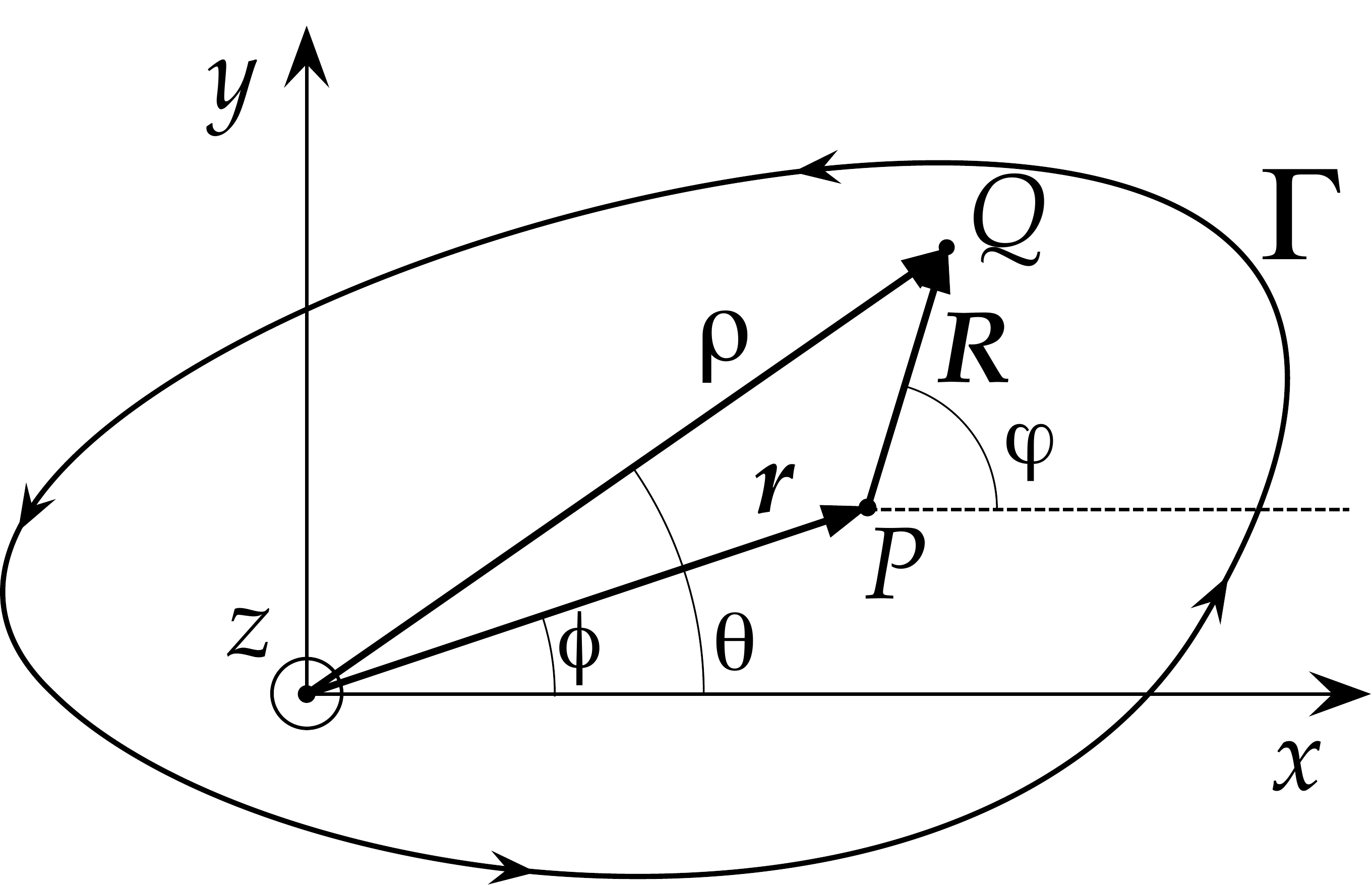}}
\caption{Describing the sharp-edge aperture from different reference frames.}
\label{Fig:FresnelIntegra}
\end{figure}

In Fig.~\ref{Fig:FresnelIntegra} the different viewpoints are shown. The $z$-axis will be supposed to be the mean propagation direction of the incident beam $\psi_0(\vettoreErre)$. 
Under very general hypotheses Fresnel's integral, written both in the forms~(\ref{Eq:FresnelIntegral}) and~(\ref{Eq:FresnelIntegralBis}), can  be reduced, in principle, to a 
(1D) \emph{contour} integral defined onto the boundary $\Gamma$. 
Hannay first noted that, for plane-wave illumination (i.e.,  $\psi_0=1$), the conversion of the Fresnel integral into Eq.~(\ref{Eq:FresnelIntegralBis}) can be 
achieved in a trivial way simply by expressing the integration variable $\vettoreR$ through its polar coordinates 
(see Fig.~\ref{Fig:FresnelIntegra}), in such a way that~\cite{Hannay/2000,Borghi/2015,Borghi/2016}
\begin{equation}
\label{Eq:Examples.2}
\begin{array}{l}
\displaystyle
\psi(\vettoreErre;U)\,=\,
\dfrac 1{2\pi}
\oint_{\Gamma}\,
\left[1\,-\,\exp\left(\ii \dfrac U2 R^2\right)\right]\,\dd\varphi\,.
\end{array}
\end{equation}

For a typical incident disturbance $\psi_0$, the surface-to-line conversion could be achieved by interpreting the integrals into Eqs.~(\ref{Eq:FresnelIntegral}) and~(\ref{Eq:FresnelIntegralBis}) as fluxes of suitable divergence-free transverse vectorial fields, say 
$\vettoref(\boldsymbol{\rho})=f(\boldsymbol{\rho})\kversore$ and $\vettoreF(\vettoreR)=F(\vettoreR)\kversore$, respectively,  where
\begin{equation}
\label{Eq:FresnelIntegral.3}
\left\{
\begin{array}{l}
\displaystyle
f(\boldsymbol{\rho})\,=\,
-\frac{\mathrm{i}U}{2\pi}\,
\psi_0(\boldsymbol{\rho})\,
\exp\left(\frac{\mathrm{i}U}{2}\,|\boldsymbol{\rho}-\vettoreErre|^2\right)\,,\\
\\
\displaystyle
F(\vettoreR)\,=\,
-\frac{\mathrm{i}U}{2\pi}\,
\psi_0(\vettoreR\,+\,\vettoreErre)\,
\exp\left(\frac{\mathrm{i}U}{2}\,R^2\right)\,,
\end{array}
\right.
\end{equation}
and where $\kversore$ denotes the unit vector of the $z$-axis.
In this way Eqs.~(\ref{Eq:FresnelIntegral}) and~(\ref{Eq:FresnelIntegralBis}) can formally be recast as follows:
\begin{equation}
\label{Eq:FresnelIntegral.2}
\begin{array}{l}
\displaystyle
\psi\,=\,\int_{\boldsymbol{\mathcal{A}}}\,
\vettoreF(\vettoreR)\,\cdot\,\kversore\,\,\,\mathrm{d}^2R\,=\,
\int_{\boldsymbol{\mathcal{A}}}\,
\vettoref(\boldsymbol{\rho})\,\cdot\,\kversore\,\,\,\mathrm{d}^2\rho\,,
\end{array}
\end{equation}
where, in order to alleviate notation complexity, the explicit dependence of $\psi$ on $U$ and $\vettoreErre$ will be tacitly assumed henceforth. 
%
%
%
%
%
The following mathematical theorem will then play a major role in the subsequent analysis: 
\begin{quotation}
{\em Consider the transverse vectorial field $\vettoreF(\vettoreR)=F(\vettoreR)\,\kversore$. Then, under very general hypotheses about the 
scalar field $F(\vettoreR)$, it is  possible to set $\vettoreF=\Nabla\,\times\,\vettoreG$, where 
\begin{equation}
\label{Eq:Asvestas.1}
\begin{array}{l}
\displaystyle
\vettoreG(\vettoreR)\,=\,\kversore\,\times\,\vettoreR\,A(\vettoreR)\,,
\end{array}
\end{equation}
and
\begin{equation}
\label{Eq:Asvestas.1.0}
\begin{array}{l}
\displaystyle
A(\vettoreR)\,=\,\int_0^1\,F(\tau\,\vettoreR)\,\tau\,\dd\,\tau\,.
\end{array}
\end{equation}
}
\end{quotation}
The same holds by formally letting $F\,\to\,f$, $\vettoreR\,\to\,\boldsymbol{\rho}$, $A\,\to\,a$.
This theorem follows from a more general theorem about vector potential representation of divergence-free vectorial fields
in the three-dimensional Euclidean space. Readers are encouraged to go through~\cite{Gordon/1975,Asvestas/1985a,Asvestas/1985b,Forbes/Asatryan/1998} 
to appreciate the simplicity and the elegance of the proof, first due to Poincar\'e~\cite{Gordon/1975}, which, as Forbes and Asatryan pointed out in their work,
``deserves to be in all the handbooks and texts, but it is not''~\cite{Forbes/Asatryan/1998}. A notable exception is the beautiful textbook by Wilfred Kaplan~\cite{Kaplan/2003}.

\begin{figure}[!ht]
\centerline{\includegraphics[width=5cm,angle=0]{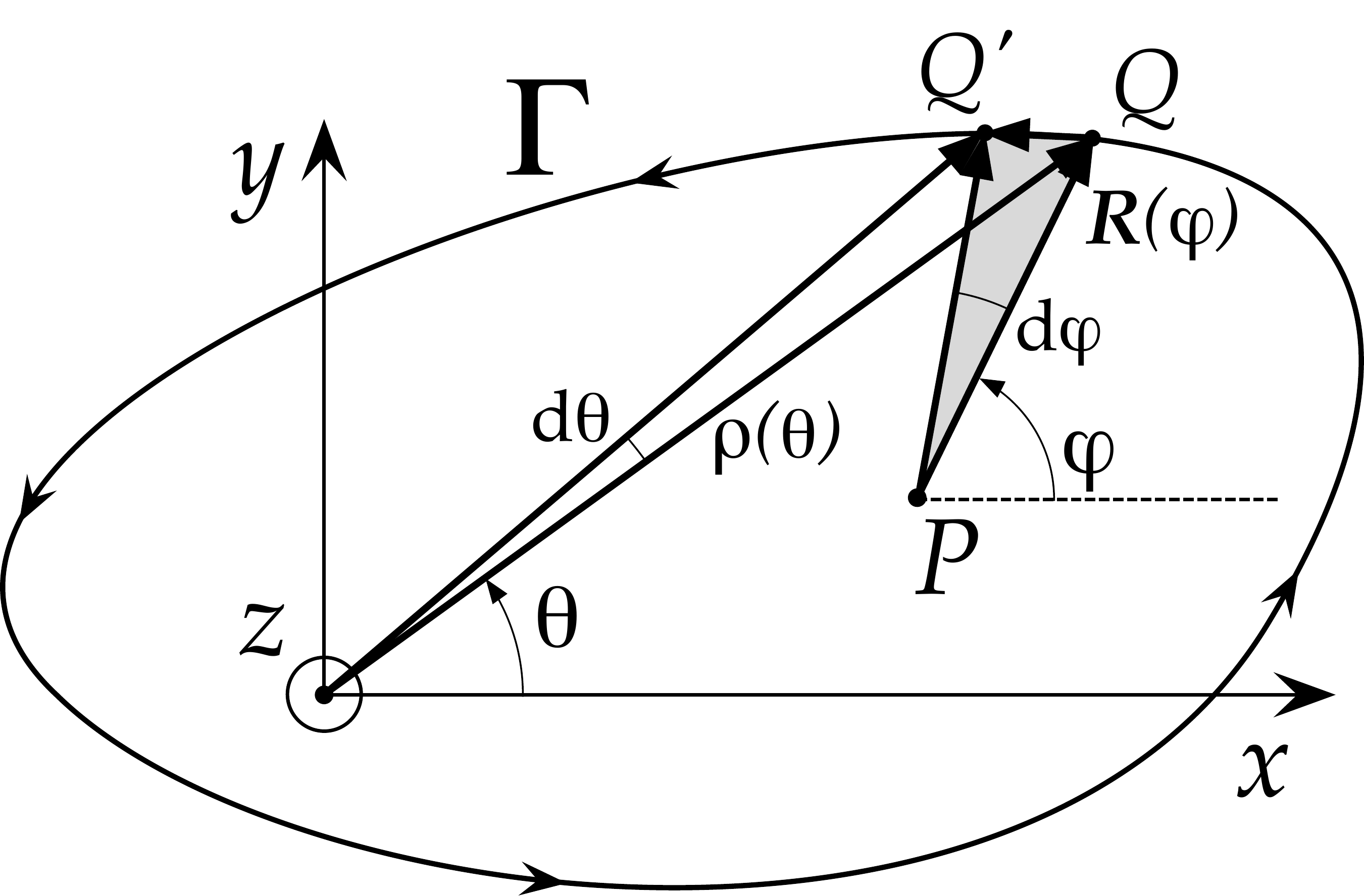}}
\caption{The geometrical meaning of $\dd\varphi=\vettoreR\,\times\,\mathrm{d}\vettoreR\,\cdot\,\kversore/R^2$ and
of $\dd\theta=\vettoreRho\,\times\,\mathrm{d}\vettoreRho\,\cdot\,\kversore/\rho^2$.}
\label{Fig:Examples.1}
\end{figure}

Now, using $F(\vettoreR)$ or  $f(\boldsymbol{\rho})$ will allow different interpretation schemes for the paraxial field $\psi$ to be given. Before doing this, 
it is worth checking Eq.~(\ref{Eq:Examples.2}). 
To this end, it is sufficient to employ the potential vector 
given  by Eq.~(\ref{Eq:Asvestas.1}), together with 
Stokes' theorem, which gives at once
\begin{equation}
\label{Eq:FresnelIntegral.4}
\begin{array}{l}
\displaystyle
\psi\,=\,\oint_{\Gamma}\,
A(\vettoreR)\,\kversore\,\times\,\vettoreR\,\cdot\,\mathrm{d}\vettoreR\,=\,
\oint_{\Gamma}\,
A(\vettoreR)\,\vettoreR\,\times\,\mathrm{d}\vettoreR\,\cdot\,\kversore\,.
\end{array}
\end{equation}
Then, on further  introducing the infinitesimal angle $\dd\varphi=\vettoreR\,\times\,\mathrm{d}\vettoreR\,\cdot\,\kversore/R^2$ sketched in Fig.~\ref{Fig:Examples.1}, 
Eq.~(\ref{Eq:FresnelIntegral.4}) can be recast as 
\begin{equation}
\label{Eq:FresnelIntegral.5.1}
\begin{array}{l}
\displaystyle
{\psi\,=\,\oint_{\Gamma}\,
R^2\,A(\vettoreR)\,\dd\varphi}\,,
\end{array}
\end{equation}
while, on setting  $\psi_0=1$ into the second row of Eq.~(\ref{Eq:FresnelIntegral.3}), Eq.~(\ref{Eq:Asvestas.1.0}) gives
\begin{equation}
\label{Eq:Examples.1}
\begin{array}{l}
\displaystyle
A(\vettoreR)\,=\,-\dfrac{\ii U}{2\pi}\,\int_0^1\,\exp\left(\ii \dfrac U2 R^2\tau^2\right)\,\tau\,\dd \tau\,=\,\\
\\
\qquad\quad\,=\,\dfrac{1\,-\,\exp\left(\ii \dfrac U2 R^2\right)}{2\pi R^2}\,.
\end{array}
\end{equation}
On substituting from  Eq.~(\ref{Eq:Examples.1}) into Eq.~(\ref{Eq:FresnelIntegral.5.1}), trivial algebra leads to Eq.~(\ref{Eq:Examples.2}). 

The use of $\boldsymbol{\rho}$ instead of $\vettoreR$ as integration variable, thus of
$\vettoref$ in place of $\vettoreF$ into Eq.~(\ref{Eq:FresnelIntegral.2}), provides new and interesting results.
To this aim, it is sufficient to recast the first of Eq.~(\ref{Eq:FresnelIntegral.3}) through the identity
$|\boldsymbol{\rho}-\vettoreErre|^2=\rho^2+r^2-2\boldsymbol{\rho}\cdot\vettoreErre$, so that the following integral representation of $\psi$
is obtained:
%
%
%
%
%
%
%
\begin{equation}
\label{Eq:Examples.2.1}
\begin{array}{l}
\displaystyle
\psi(\vettoreErre;U)\,=\,
-\dfrac{\ii U}{2\pi}\,\exp\left(\ii\dfrac U2 r^2\right)
\oint_{\Gamma}\,
a(\boldsymbol{\rho})\,\boldsymbol{\rho}\,\times\,\mathrm{d}\boldsymbol{\rho}\,\cdot\,\kversore\,,
\end{array}
\end{equation}
where now
\begin{equation}
\label{Eq:Examples.2.2}
\begin{array}{l}
\displaystyle
a(\boldsymbol{\rho})\,=\,
\int_0^1\,\dd \tau\,\tau\,\,\psi_0(\tau\,\boldsymbol{\rho})\,\exp\left(\ii\dfrac U2\,\left[\tau^2\rho^2-2\tau\boldsymbol{\rho}\cdot\vettoreErre\right]\right)\,.
\end{array}
\end{equation}
%
%
%
%
%
%
%
%

Before continuing, it is worth recalling  an important aspect of the 
present formulation. If the initial field $\psi_0(\vettoreErre)$ does allow the analytical exact evaluation of the 
integral~(\ref{Eq:Examples.2.2}), then the diffracted field $\psi$ will be expressed by a single contour integral. 
If not, the propagated field $\psi$ will unavoidably be expressed through a double integral, whose numerical 
evaluation should  expect to be easier than that of the original Fresnel's integrals~(\ref{Eq:FresnelIntegral}) 
and~(\ref{Eq:FresnelIntegralBis}). 
However, if $\Gamma$ were  circular (unit radius), Eqs.~(\ref{Eq:Examples.2.1}) and~(\ref{Eq:Examples.2.2}) 
would be equivalent to compute Fresnel's integral~(\ref{Eq:FresnelIntegral}) via polar coordinates $(\rho,\theta)$,  being $\tau\equiv\rho$ . 
The computational novelty of Eqs.~(\ref{Eq:Examples.2.1}) and~(\ref{Eq:Examples.2.2}) can then be unveiled by exploring sharp-edge diffraction from 
\emph{noncircular} apertures. 
In fact, differently from Fresnel, whose integration domain coincides with the 
aperture $\CalA$, Eqs.~(\ref{Eq:Examples.2.1}) and~(\ref{Eq:Examples.2.2}) imply that, regardless the aperture shape, 
the new double integral representation of $\psi$ will be \emph{always} defined onto the Cartesian product between the $\tau$ integration interval 
$[0,1]$ and the (finite) integration interval related to the parametrization of the boundary $\Gamma$. As we shall see, this allows standard 
Montecarlo integration techniques on hypercubes to be efficiently employable.

\begin{figure}[!ht]
\centerline{\includegraphics[width=4cm,angle=-0]{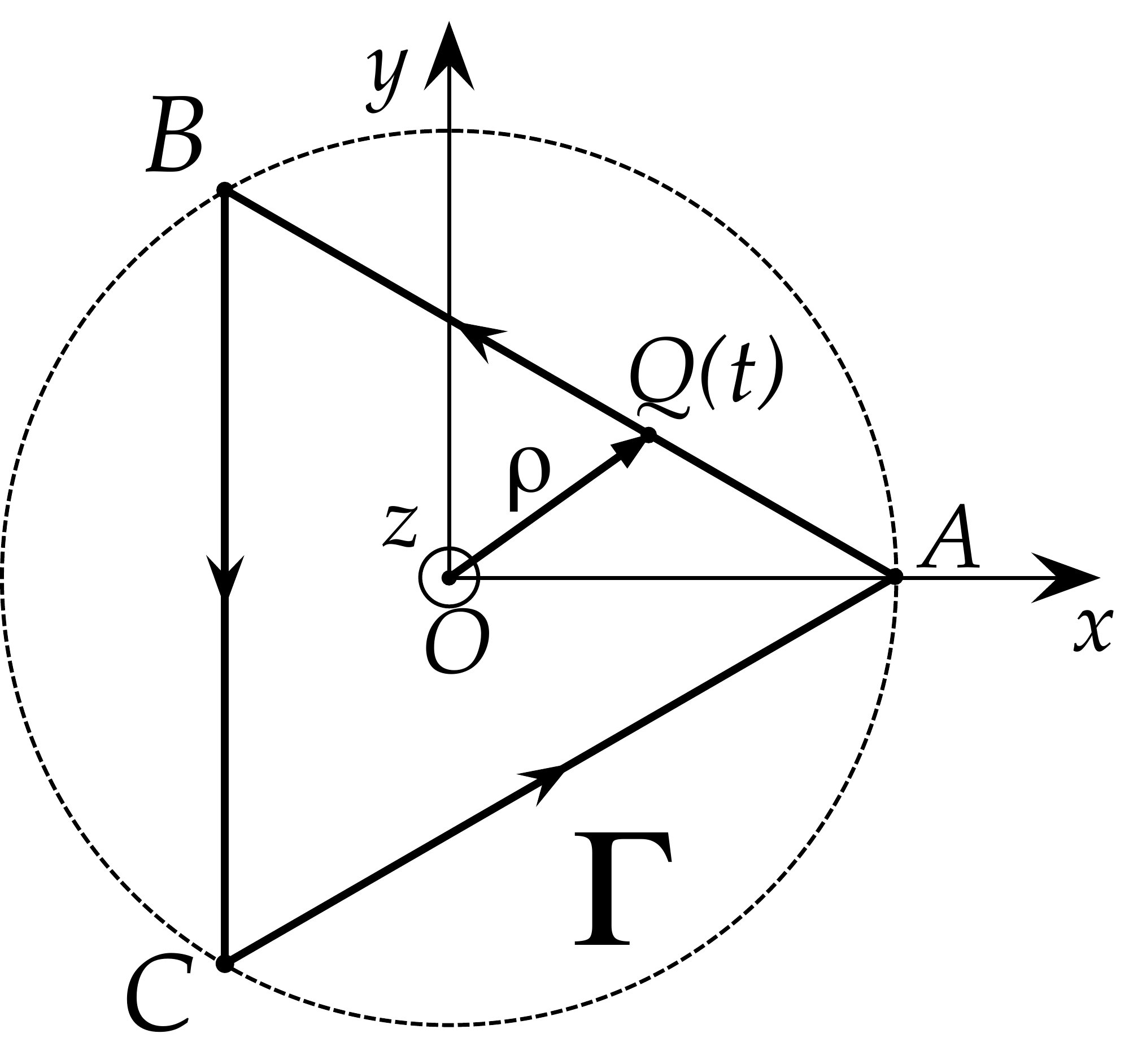}}
\caption{The equilateral triangular diffracting sharp-edge aperture.}
\label{Fig:Triangle}
\end{figure}

A couple of examples will now be illustrated which, for simplicity, involve the same aperture, namely the equilateral triangle  $ABC$ shown 
in  Fig.~\ref{Fig:Triangle}. The characteristic length will then be identified by the radius $\overline{OA}$ of the 
circumscribed circle. Accordingly, in the following it will be set $\overline{AB}=\sqrt 3$.
Each side of the boundary $\Gamma$ will be parametrized according to the most natural choice. So, the parametrization of the side $AB$ reads (see Fig.~\ref{Fig:Triangle})
\begin{equation}
\label{Eq:Examples.2.2.1}
\begin{array}{l}
\displaystyle
\boldsymbol{\rho}(t)\,=\,\overrightarrow{OQ(t)}\,=\,\overrightarrow{OA}\,+\,\overrightarrow{AB}\,t\,,\qquad\qquad\qquad 0 \le t \le 1\,,
\end{array}
\end{equation}
and similarly for the other two sides. In this way, it is trivial to prove that  $\boldsymbol{\rho}\,\times\,\mathrm{d}\boldsymbol{\rho}\,\cdot\,\kversore={\sqrt 3}/2\,\dd t$ inside Eq.~(\ref{Eq:Examples.2.1}),
what considerably simplifies the evaluation of the contour integral $\oint_\Gamma$. 

The first example deals with the effect of a sharp-edge aperture on the propagation of light beams 
carrying on vortices of given  topological charge, say $m\ge 1$.
A simple analytical model for the impinging beam is
\begin{equation}
\label{Eq:Gaussian.1}
\begin{array}{l}
\displaystyle
\psi_0(\vettoreErre)\,=\,r^m\,\exp(\ii\,m\,\phi)\,\exp(\ii \alpha r^2)\,, \quad\quad m \ge 1,\,\,\alpha \in \mathbb{C}\,,
\end{array}
\end{equation}
which represents a particular Laguerre-Gauss distribution. 
On substituting from Eq.~(\ref{Eq:Gaussian.1}) into Eq.~(\ref{Eq:Examples.2.2}) we have 
\begin{equation}
\label{Eq:Gaussian.2}
\begin{array}{l}
\displaystyle
a(\boldsymbol{\rho})\,=\,\rho^m \exp(\ii m\theta)\\
\\
\displaystyle
\times\,
\int_0^1\,\dd \tau\,\tau^{m+1}\exp\left[\ii\left(\alpha\,+\,\dfrac U2\right)\,\tau^2\rho^2\,-\,\ii\tau\,U\,\boldsymbol{\rho}\cdot\vettoreErre\right]\,.
\end{array}
\end{equation}
Now, it can be proved that the whole family of integrals:
\begin{equation}
\label{Eq:Gaussian.4}
\begin{array}{l}
\displaystyle
\mathcal{I}_n(L;\,M)\,=\,\int_0^1\,\dd\tau\,
\tau^n\,\exp[\ii (L \tau^2\,-\, M\,\tau)]\,,
\end{array}
\end{equation}
can be evaluated exactly, for \emph{any} complex $L$ and real $M$, through the following notable recursive rule:
\begin{equation}
\label{Eq:Gaussian.4.1}
\left\{
\begin{array}{l}
\displaystyle
\mathcal{I}_0\,=\,(-)^{1/4}\,\exp\left(\dfrac{-\ii M^2}{4L}\right)\,
\sqrt{\dfrac{\pi}{4\,L}}\,\\
\\
\times\,
\left(
\erf\left[(-)^{3/4}\,\dfrac{M-2L}{\sqrt{{4 L}}}\right]\,-\,
\erf\left[(-)^{3/4}\dfrac{M}{\sqrt{{4 L}}}\right]
\right)\,,\\
\\
\mathcal{I}_1\,=\,
\dfrac{\ii}{2L}\,\left(1\,-\,\exp[\ii(L-M)]\right)\,+\,\dfrac M{2L}\,\mathcal{I}_0\,,\\
\\
\mathcal{I}_n\,=\,
\dfrac M{2L}\,\mathcal{I}_n\,+\,\dfrac{\ii n}{2L}\,\mathcal{I}_{n-1}\,-\,
\dfrac{\ii}{2L}\,\exp[\ii(L-M)]\,,\quad n>1\,,
\end{array}
\right.
\end{equation}
which is one of the most relevant results of the present paper.
Equation~(\ref{Eq:Gaussian.4.1}) provides an important generalization of the results recently found in~\cite{Borghi/2019,Borghi/2020} to sharp-edge diffraction 
of Gaussian beams~\cite{Worku/Gross/2019}. 

On using Eq.~(\ref{Eq:Gaussian.2}) into Eq.~(\ref{Eq:Examples.2.1}) and on recalling the 
 above described triangle parametrization, the diffracted wavefield $\psi(\vettoreErre;U)$ can be written (apart from 
unessential amplitude and overall phase factors) as
\begin{equation}
\label{Eq:Gaussian.5}
\begin{array}{l}
\displaystyle
\psi(\vettoreErre;U)\,\propto\,
\sum_{j=1}^3\,
\int_0^1\,\mathrm{d} t\,
(\xi_j+\ii\eta_j)^m\,
\mathcal{I}_{m+1}\left[\alpha+\dfrac U2;\,U (\xi_j x+\eta_j y)\right]\,,
\end{array}
\end{equation}
where $\boldsymbol{\rho}_j(t)=(\xi_j(t),\eta_j(t))$ defines the parametrization of the $j$th triangle side.

\begin{figure}[!ht]
\begin{minipage}[t]{9cm}
{\includegraphics[width=4cm,angle=-0]{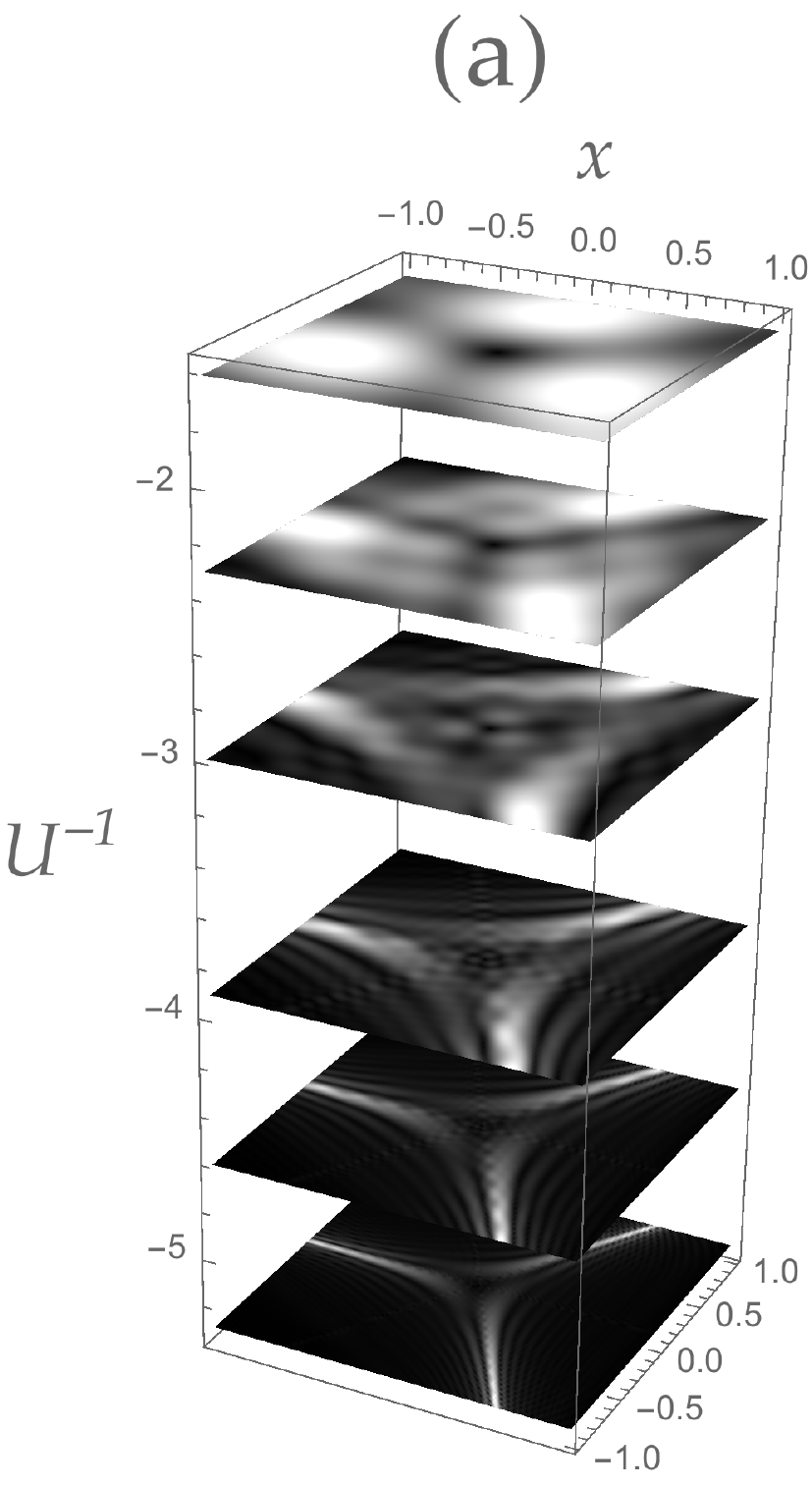}}
\hspace*{0.0cm}
{\includegraphics[width=4cm,angle=-0]{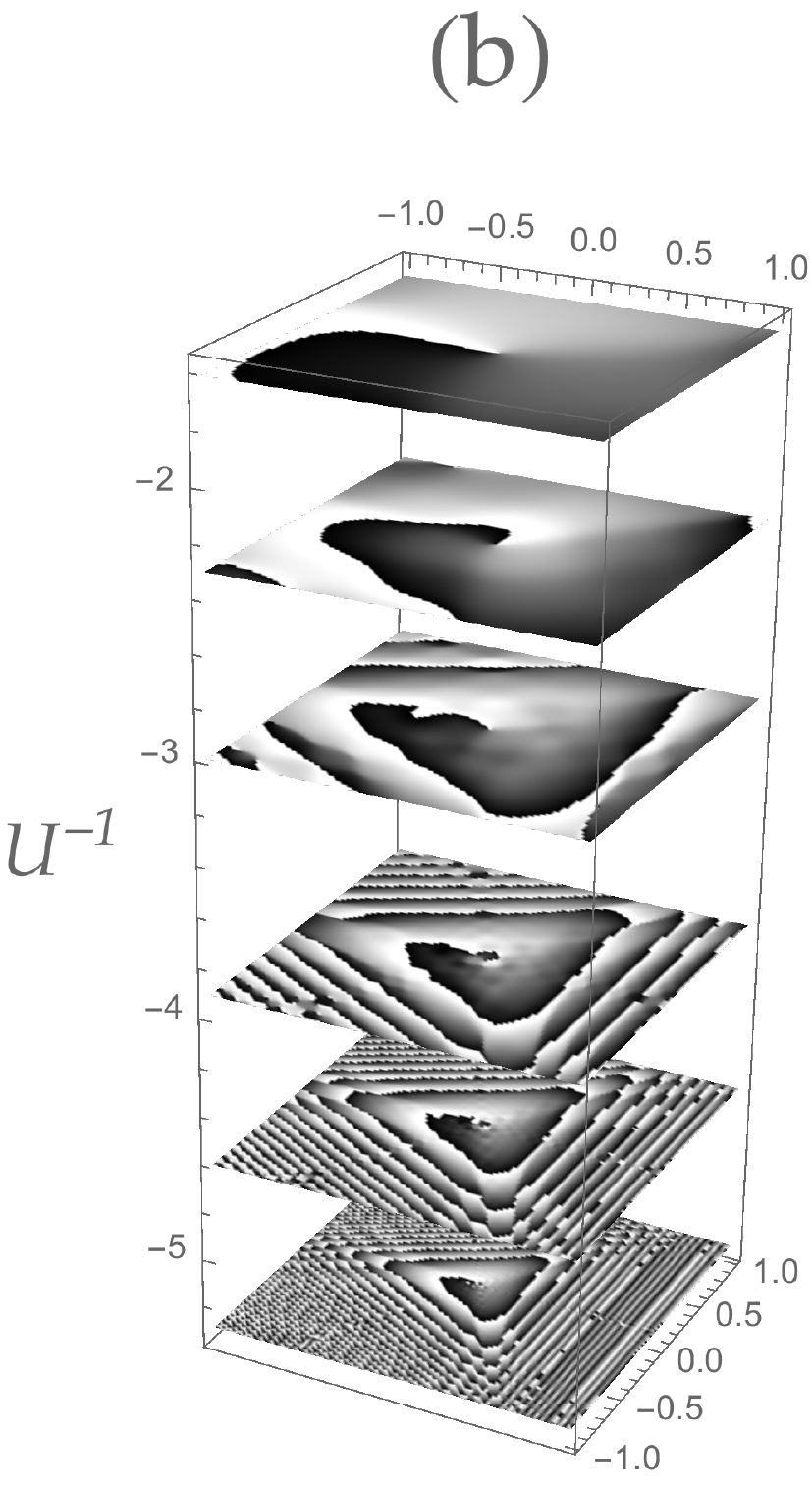}}
\end{minipage}
%
\caption{2D maps of the modulus (a) and the phase (b) of the diffracted field $\psi(\vettoreErre;U)$, evaluated via Eq.~(\ref{Eq:Gaussian.5}), are plotted for $U=5,\,10,\,20,\,50,\,100,\,200$.}
\label{Fig:VortexDiffraction}
\end{figure}

The results of our numerical experiment are shown in Fig.~\ref{Fig:VortexDiffraction}.
A collimated LG beam carrying on vortex with unitary topological charge impinges onto the triangular shape of 
Fig.~\ref{Fig:Triangle}. The spot-size of the incident beam will be set to $w_0=\pi\,a$, i.e., larger enough than the 
aperture size to simulate a plane wave carrying on a unitary charge vortex, as suggested in~\cite{Rocha/Amaral/Fonseca/Jesus-Silva/2019}. In Fig.~\ref{Fig:VortexDiffraction}, 2D maps of the modulus (a) and the phase (b) of the diffracted field $\psi(\vettoreErre;U)$, numerically evaluated via Eq.~(\ref{Eq:Gaussian.5}), are 
plotted for $U=5,\,10,\,20,\,50,\,100,\,200$ (note that the $U$ scale is logarithmic). 

From the figure it is possible in particular to appreciate the evolution of the topological complexity
of the 2D field distribution from the near ($U=200$, bottom) to the far ($U=5$, top) zone. 
The possibility of modeling near-field diffraction of vortex beams in such a simple way should be positively
acknowledged as a further powerful investigation tool to explore light's orbital angular momentum. 
In fact, since the pioneering work~\cite{Hickmann/Fonseca/Soares/Chavez-Cerda/2010},
experimental and theoretical investigations have mainly been focused on Fraunhofer diffraction, a
considerably easier scenario to be numerically simulated with respect to Fresnel~\cite{Stahl/Gbur/2016,Rocha/Amaral/Fonseca/Jesus-Silva/2019,Kun-Rui/Jian-Nan/Zhi-Kun/2021}.

%
%
%
%

The second, nearly iconic, example of application  of Eqs.~(\ref{Eq:Examples.2.1}) and~(\ref{Eq:Examples.2.2}) will now be illustrated. 
The following quotation well describes the experimental situation~\cite{Berry/Nye/Wright/1979}:
\begin{quotation}
{\em\noindent A hole whose shape approximated an equilateral triangle of side $\ell=$2.6~mm was cut in adhesive tape 
stuck on to the horizontal surface of a glass microscope slide. A water droplet was allows to fall on to the slide, where 
it formed a thin lens. This lens was illuminated from below with a parallel beam of laser light (wavelength 
$\lambda$= 633~nm) broadened so as to fill the aperture of the lens. After refraction the focused light formed an 
elliptic umbilic diffraction catastrophe a few centimetres above the lens.
}
\end{quotation}
The experiment was aimed at producing what is known to be
an \emph{elliptic umbilic diffraction catastrophe}. Diffraction catastrophes are mathematical bricks with which, 
at the end of seventies, John Nye and Michael Berry founded the so-called \emph{Catastrophe Optics}: a new, modern  
theoretical framework aimed at studying the so-called ``natural focusing'' of light~\cite{Nye/1999,Berry/Upstill/1980}. 
CO's description of focused wavefields is built up starting from light skeletons of bright caustics which are decorated,
at the wavelength scale, by characteristic diffraction patterns organized according to a precise hierarchy~\cite{Berry/Upstill/1980}. The elliptic umbilic is one of them.

Some of the experimental results presented in~\cite{Berry/Nye/Wright/1979} will now be reproduced thanks to a straightforward
implementation of the general method here developed. To this aim, the radius of the circumscribed circle to the triangle $a=\ell/\sqrt 3$ is again introduced, while the refractive index of the droplet will be set to $n \simeq 4/3$. 
In~\cite{Berry/Nye/Wright/1979}  a simple and effective physical model of the droplet profile, based on the combined action of  
surface tension and gravity, was developed (see also~\cite{Nye/1986}). 
In particular, on denoting $h(\vettoreErre)$ the height of the droplet upper surface below the plane $z = 0$, we have
\begin{equation}
\label{Eq:BHW.2.1.1}
\begin{array}{l}
\displaystyle
h(\vettoreErre)\,=\,\left\{
\begin{array}{lr}
H\,(1\,-\,3\,{r^2}\,+\, 2\,{x^3\,-\,6\,\,xy^2})\,,&\vettoreErre\,\in\,\CalA\,,\\
\\
0\,&\vettoreErre\,\notin\,\CalA\,,
\end{array}
\right.
\end{array}
\end{equation}
where $H\,\simeq\,0.1$~mm, denotes the maximum height of the droplet.
\begin{figure}[!ht]
\centerline{\includegraphics[width=5cm,angle=-0]{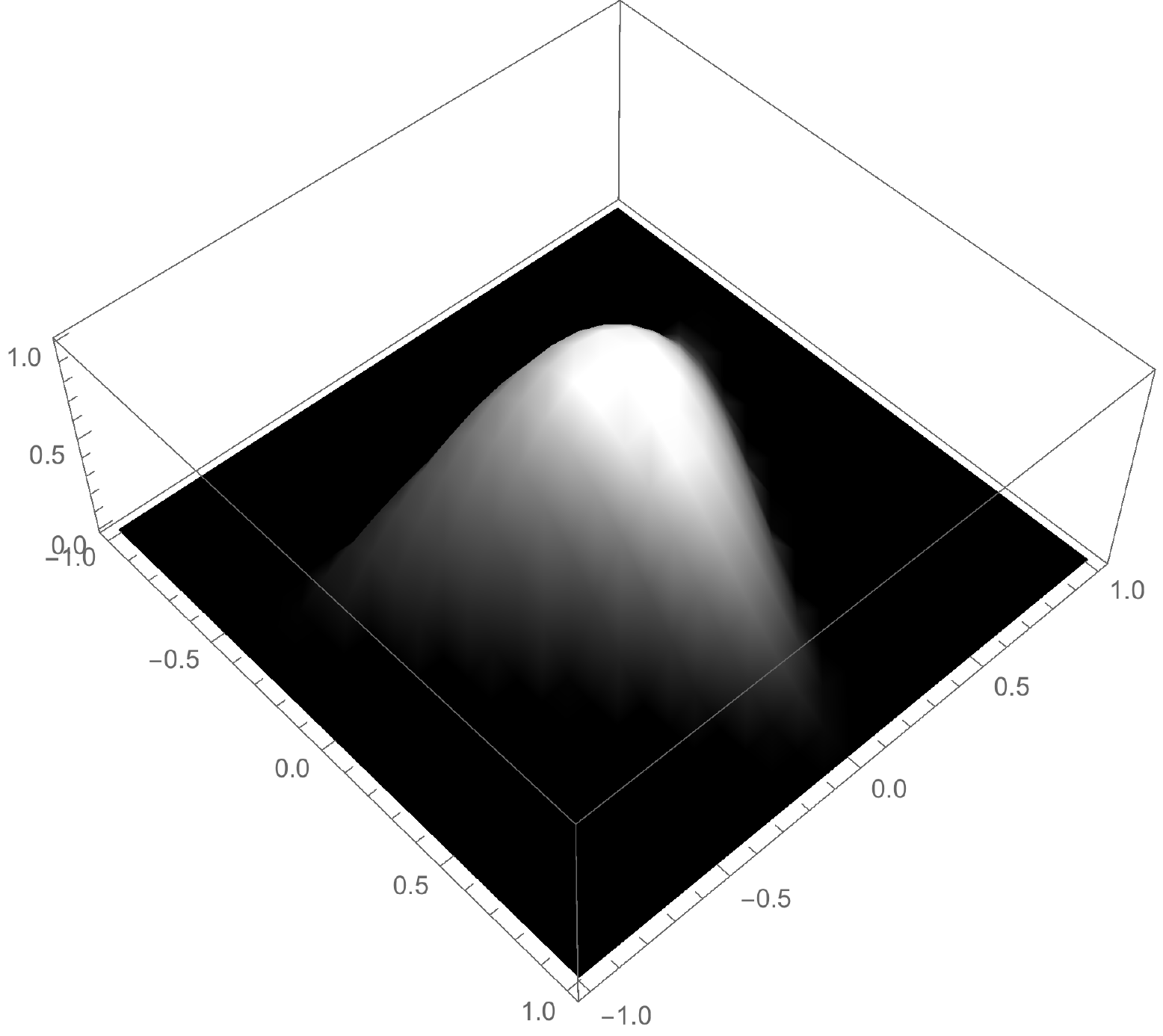}}
\caption{3D visualization of the droplet normalized profile $h(\vettoreErre)/H$ given in Eq.~(\ref{Eq:BHW.2.1.1}).}
\label{Fig:Droplet3D}
\end{figure}

A 3D visualization of the normalized profile $h/H$ is sketched in Fig.~\ref{Fig:Droplet3D}. 
Once the droplet is illuminated by the laser beam, the diffracted field can be obtained (up to  unessential overall phase factors) 
by evaluating the Fresnel integral into Eq.~(\ref{Eq:FresnelIntegral}) with $\psi_0$ given by
\begin{equation}
\label{Eq:BHW.3}
\begin{array}{l}
\displaystyle
\psi_0(\vettoreErre)\,=\,\exp[-\ii \alpha\, (3\,{x^2}\,+\,3\,{y^2}\,-\,2\,{x^3\,+6\,\,xy^2})]\,,
\end{array}
\end{equation}
where now $\alpha=k\,(n-1)\,H\,\simeq\,330.868\ldots$
The values of $U$ will be chosen according to the prescriptions described in~\cite{Berry/Nye/Wright/1979}. 
In particular, the droplet lens focus is located at a distance from the aperture plane equal to 
$\ell^2/(18 H (n-1))\,\simeq\,11.2$~mm, 
corresponding to $U\,=\,6\alpha\simeq2\,\times\,10^3$. The subsequent simulations will then be carried out in the neighborhood of such a 
value. 

The {nonregular} shape of $\CalA$, together with the nonsmall values of $U$ (of the order of thousands), would make the evaluation of the 2D Fresnel 
integral~(\ref{Eq:FresnelIntegral}) a challenging numerical task.
In~\cite{Berry/Nye/Wright/1979} such technical difficulties were partially circumvented by suitably extending the integration domain $\CalA$
to cover the whole Euclidean plane $\mathbb{R}^2$, thus neglecting the edge wave contribution. In other words,
$\psi(\vettoreErre;U)$ was  approximated through an elliptic umbilic diffraction catastrophe, whose evaluation can be  
achieved for instance by using suitable asymptotic techniques.
In the following, Eqs.~(\ref{Eq:Examples.2.1}) and~(\ref{Eq:Examples.2.2}) will be employed to reproduce some of the experimental 
results shown in Fig.~2 of~\cite{Berry/Nye/Wright/1979}, without any approximations but the paraxial one. 
To help the comparison, 
Fresnel's number $U$ will be recast as follows:
\begin{equation}
\label{Eq:BHW.3.1}
\begin{array}{l}
\displaystyle
U\,=\,6\,\alpha\,-\,2\,(2\,\alpha)^{2/3}\,\zeta\,,
\end{array}
\end{equation}
where the symbol $\zeta$ denotes a dimensionless, normalized abscissa whose origin is located at the above defined focal plane. 
In particular, the definition~(\ref{Eq:BHW.3.1}) has been chosen in order for $\zeta$ to  coincide with the parameter employed 
in~\cite{Berry/Nye/Wright/1979} to individuate the position of the observation plane during their experiment.
On substituting from Eq.~(\ref{Eq:BHW.3}) into Eq.~(\ref{Eq:Examples.2.2}) and then into Eq.~(\ref{Eq:Examples.2.1}), and on taking
Eq.~(\ref{Eq:BHW.3.1}) into account, the diffracted wavefield $\psi(\vettoreErre;U)$ can formally be written 
(apart from unessential amplitude and overall phase factors) as follows: 
\begin{equation}
\label{Eq:BHW.3.1.1}
\begin{array}{l}
\displaystyle
\psi\,\propto\,
\sum_{j=1}^3
\int_0^1
\int_0^1
\dd t\,\dd \tau\,\tau\,
\exp\left[2\,\ii \alpha\,\tau^3 ({\xi_j^3\,-\,3\,\xi_j\eta_j^2})\right]\\
\qquad\qquad\qquad\times\exp\left[-\ii\,(2\,\alpha)^{2/3}\,\zeta\,\tau^2\rho_j^2\,-\,\ii\,U\,\tau\boldsymbol{\rho}_j\cdot\vettoreErre\right]\,,
\end{array}
\end{equation}
with the same symbol meaning as for Eq.~(\ref{Eq:Gaussian.5}).
Each double integral into Eq.~(\ref{Eq:BHW.3.1.1}) will be evaluated via standard Montecarlo techniques. To give an idea, 
all subsequent figures have been produced by using  \emph{Wolfram Mathematica} native routine \texttt{NIntegrate} with the following options: 
\begin{quotation}
\noindent\texttt{Method} $\to$ \texttt{"AdaptiveMonteCarlo"}\\
\\
\texttt{"MaxPoints"} $\to10^5$
\end{quotation}
%

\begin{figure}[!ht]
\centerline{\includegraphics[width=4cm,angle=-0]{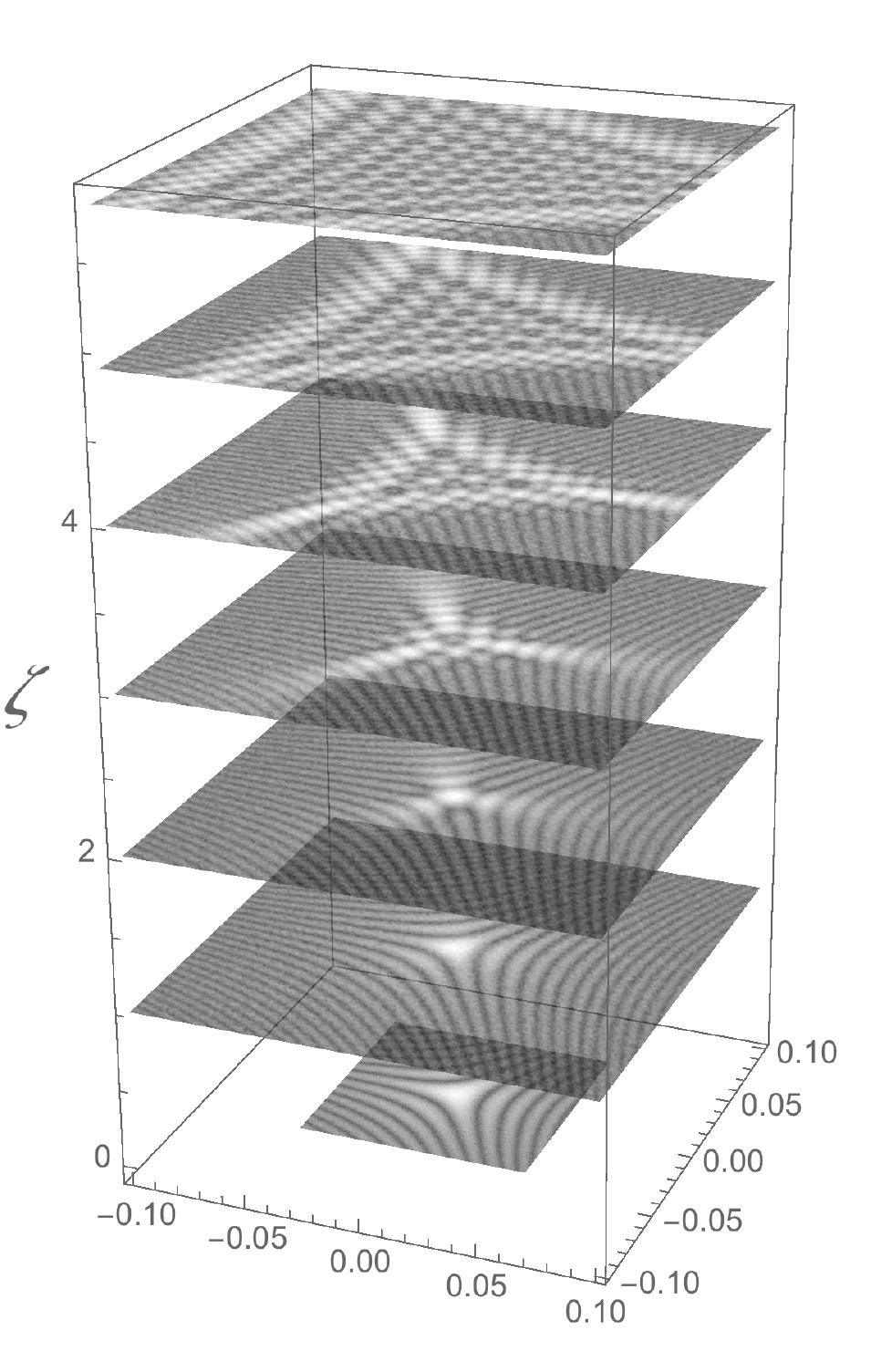}}
\caption{2D maps of the optical intensity distribution of the field focused by the droplet lens of Fig.~\ref{Fig:Droplet3D}, evaluated via Eq.~(\ref{Eq:Gaussian.5}), 
are plotted for $\zeta=0,\,1,\,2,\,3,\,4,\,4.9,\,5.81$.}
\label{Fig:BerryNyeWright}
\end{figure}
\begin{figure}[!ht]
\centerline{\includegraphics[width=8cm,angle=-0]{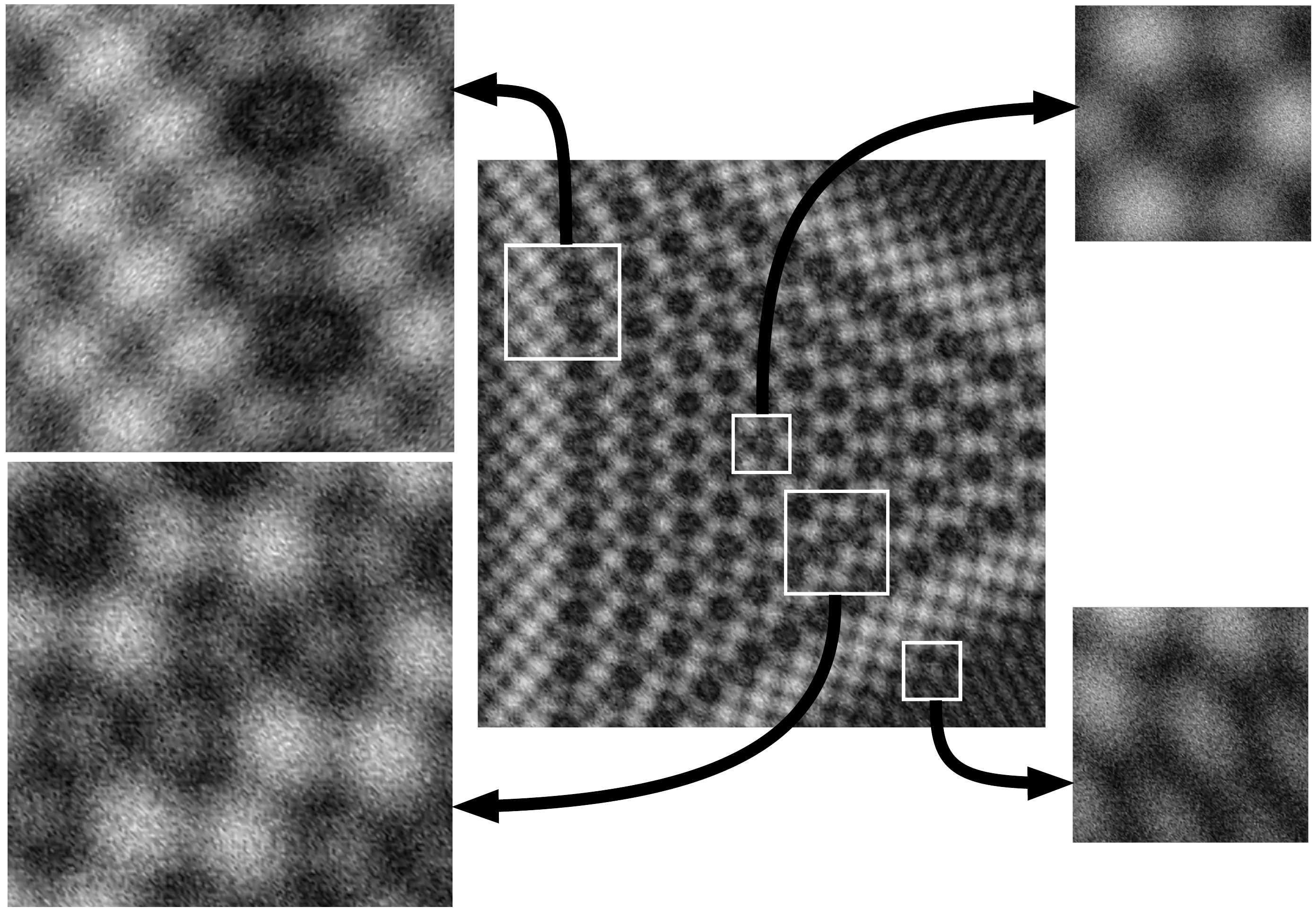}}
\caption{Blowups of the upper slice of Fig.~\ref{Fig:BerryNyeWright}, corresponding to 
$\zeta=5.81$.}
\label{Fig:BerryNyeWright.2}
\end{figure}

In Fig.~\ref{Fig:BerryNyeWright}, 2D maps of the transverse intensity distribution $|\psi(\vettoreErre;U)|^2$
are shown for some values of the dimensionless absicssa $\zeta$. They are aimed at reproducing the 
experimental pictures reported in Fig. 2 of Ref.~\cite{Berry/Nye/Wright/1979}, precisely
Figs.~2(a),~2(b),~2(c),~2(d),~2(g),~2(h),~and 2(i),  corresponding to  
$\zeta=0$ (a), 1 (b), 2 (c), 3 (d), 4 (g), 4.9 (h), and 5.81 (i), respectively.
Each slice of Fig.~\ref{Fig:BerryNyeWright} contains a $200\,\times\,200$ matrix of intensity values.
In Fig.~\ref{Fig:BerryNyeWright.2} some blowups of the upper slice of Fig.~\ref{Fig:BerryNyeWright}, which corresponds to 
$\zeta=5.81$, are also shown.  The figure appears to be slightly noisy, due to Montecarlo.

Nevertheless, the visual agreement between Figs.~\ref{Fig:BerryNyeWright} and~\ref{Fig:BerryNyeWright.2}  with the experimental results presented in 
Fig.~2 of~\cite{Berry/Nye/Wright/1979} is excellent.

\section{Conclusions}
\label{Sec:Conclu}

Fresnel's diffraction theory represents a milestone of classical optics
since more than two centuries. However, the numerical evaluation of 2D diffraction integrals remains a challenging
task,  due to the highly oscillatory behavior of the integrands and the shape of the integration domains. 
For plane and/or spherical waves, a reduction of Fresnel's integral to 1D, singular-free contour phase integrals is
always guaranteed as far as sharp-edge diffraction is concerned. 
Such a formulation, which has been developed during the last decade in the light of catastrophe optics, revealed to be
an unorthodox,  very interesting point of view from which diffraction phenomena can be explored.

In the present paper a further step toward a general paraxial sharp-edge diffraction theory dealing with, in principle,  arbitrary wavefields
impinging onto arbitrarily shaped planar apertures, has been proposed. By using Poincar\'e vector potential construction, Fresnel's integral has 
been converted into a contour integral over the aperture rim. In this way, it has been proved that sharp-edge diffraction of a whole subclass of Laguerre-Gauss 
beams carrying on vortices of arbitrarily high topological charges  can numerically be dealt only with 1D integrals. 
When the analytical conversion to a single contour integral is no longer possible, a new double integral representation of 
the diffracted wavefield, suitable for Montecarlo integration, has been derived. It was tested on an iconic example of natural focusing of light 
by water droplets, with extremely promising results.

All numerical simulations presented in the paper have deliberately been
carried out with a ``low profile'' approach. The computing machine employed to generate
all figures was a commercial laptop equipped with  a 3~GHz Intel Core i7 processor and 16~GB RAM.
Moreover, all numerical integrations (both 1D and 2D) presented have been performed via
standard native \emph{Mathematica} routines. This should convince our readers about the
feasibility, the implementation easiness, and the numerical effectiveness  of the proposed method also when
``extreme''  scenarios are dealt with.

Future studies are in progress. Among these, the application to cascaded diffraction in optical systems is one of  the most relevant, as witnessed by the recent 
literature~\cite{Mout/Flesch/Wick/Bociort/Petschulat/Urbach/2018,Gross/2020a,Gross/2020b}. When diffraction by a sequence of sharp-edge apertures has to be tackled, 
the approach here developed is expected to be highly promising. In particular, the emerging wavefield $\psi$ would naturally be represented in terms of multiple integrals defined onto 
hypercubes, a perfect scenario for effectively employing Montecarlo-based computational techniques.

Light scattering from ``large''  tridimensional objects is another
topic which would be worth exploring with the above computational tools.
In particular, the basic features of the scattered wavefields could, in principle, be grasped by 
replacing the scatterers by ``equivalent'' planar apertures.
Raman and Krishnan first experimentally explored this topic~\cite{Raman/Krishnan/1926}, which has continued
to receive  attention, also due to
its important astronomical implications~\cite{Roques/Moncuqet/Sicardy/1987,Melbourne/2005,Heinson/Chakrabarti/Sorensen/2014,Young/2012}.

Finally, the proposed approach could also be helpful in numerically exploring the role played by sharp-edge diffraction in the study of the fractal nature of the light, pioneered in~\cite{Berry/1979} and presently a topic of central interest in theoretical and applied optics (see for instance the review in~\cite{Korolenko/2020}).

``There is pleasure in recognizing old things from a new viewpoint,'' Richard Feynman loved to say. During the last decade,
the change of perspective offered by Catastrophe Optics in describing paraxial sharp-edge diffraction provided new light on old, nearly forgotten experiments, as well as new, 
unorthodox interpretation schemes of diffraction problems by now considered obsolete. 
We hope what has here been presented could be helpful in continuing to explore new, unexpected, and still unveiled aspects of 
classical diffraction theory.

\section*{Acknowledgements}

I wish to thank Gabriella Cincotti and Turi Maria Spinozzi for their unvaluable help during the preparation of the manuscript.

%
%

%
%
%
%

\end{document}